\documentclass{Interspeech2024}
\interspeechcameraready 
\usepackage{multicol,multirow}
\usepackage{xurl,cite}
\usepackage{hyperref}
\usepackage{xcolor, colortbl}
\usepackage{tabularx,lipsum,amssymb,amsmath,pifont,graphicx,adjustbox}
\usepackage{lipsum}
\usepackage{verbatim}
\hypersetup{
    colorlinks=true,
    linkcolor=purple,
    filecolor=magenta,      
    urlcolor=purple,
}
\definecolor{color_am_dnn}{RGB}{153, 204, 255}
\definecolor{color_am_sm}{rgb}{1, 1, 1}
\definecolor{color_wm_dnn}{RGB}{255, 204, 204}
\definecolor{color_wm_dsp}{rgb}{1, 1, 1}
\definecolor{color_tts}{RGB}{173, 235, 173}
\definecolor{color_vc}{rgb}{1, 1, 1}

\title{To what extent can ASV systems naturally defend against spoofing attacks?}
\name[affiliation={1}]{Jee-weon}{Jung$^*$}
\name[affiliation={2}]{Xin}{Wang$^*$}
\name[affiliation={3}]{Nicholas}{Evans}
\name[affiliation={1}]{Shinji}{Watanabe}
\name[affiliation={1}]{Hye-jin}{Shim}
\name[affiliation={4}]{Hemlata}{Tak}
\name[affiliation={1}]{Siddhant}{Arora}
\name[affiliation={2}]{Junichi}{Yamagishi}
\name[affiliation={5}]{Joon Son}{Chung}

\address{
  $^1$Carnegie Mellon University, USA, 
  $^2$National Institute of Informatics, Japan\thanks{$^*$Equal contribution.}\\
  $^3$EURECOM, France, 
  $^4$Pindrop, USA\\
  $^5$Korea Advanced Institute of Science and Technology, South Korea
}
\email{jeeweonj@ieee.org, wangxin@nii.ac.jp}

\newcommand{\newpara}[1]{\vspace{1pt}\noindent\textbf{#1}}

\begin{document}

\maketitle

\begin{abstract}
The current automatic speaker verification (ASV) task involves making binary decisions on two types of trials: target and non-target. However, emerging advancements in speech generation technology pose significant threats to the reliability of ASV systems. This study investigates whether ASV effortlessly acquires robustness against spoofing attacks (i.e., zero-shot capability) by systematically exploring diverse ASV systems and spoofing attacks, ranging from traditional to cutting-edge techniques. Through extensive analyses conducted on eight distinct ASV systems and 29 spoofing attack systems, we demonstrate that the evolution of ASV inherently incorporates defense mechanisms against spoofing attacks. Nevertheless, our findings also underscore that the advancement of spoofing attacks far outpaces that of ASV systems, hence necessitating further research on spoofing-robust ASV methodologies.
\end{abstract}

\section{Introduction}

Automatic Speaker Verification (ASV) systems are biometric technologies designed to confirm whether a given speech sample belongs to a claimed enrolled speaker\cite{campbell1997speaker}. These systems evaluate performance through a protocol consisting of two types of trials: target trials, where the voice in the speech sample matches the enrolled speaker, and non-target trials, where it does not. As a crucial component of an authentication system, ASV must not only differentiate between these trials but also robustly defend against spoofing attacks (also known as presentation attacks) -- efforts to deceive the system with artificially generated voices. Yet, remarkably, the integration of such critical defenses against spoofing attacks has been considered in only a fractional segment of the existing ASV systems~\cite{wu2015asvspoof,jung2022sasv}.

Meanwhile, the explosive advancements in Text-To-Speech synthesis (TTS) and Voice Conversion (VC) have heightened vulnerabilities to spoofing attacks. The accessibility of sophisticated TTS and VC systems has inadvertently equipped fraudsters with the means to generate speech data mimicking specific target individuals~\cite{GMA2023voice}. While only a subset of the latest deep-learning-based TTS and VC systems reached the perceptual quality of bonafide speech~\cite{shen2018natural, Yi2020, cooper21_ssw}, 
many systems, even those employing basic linear statistical models, have successfully fooled state-of-the-art ASV systems at the time (e.g., the i-vector PLDA \cite{dehak2010front} and the x-vector \cite{snyder2018x} systems evaluated in ASVspoof 2015 \cite{wu2015asvspoof} and 2019 challenges \cite{wang2020asvspoof}). 

\begin{figure}
    \centering
    \includegraphics[trim={0 0.4cm 0 0},width=0.82\columnwidth]{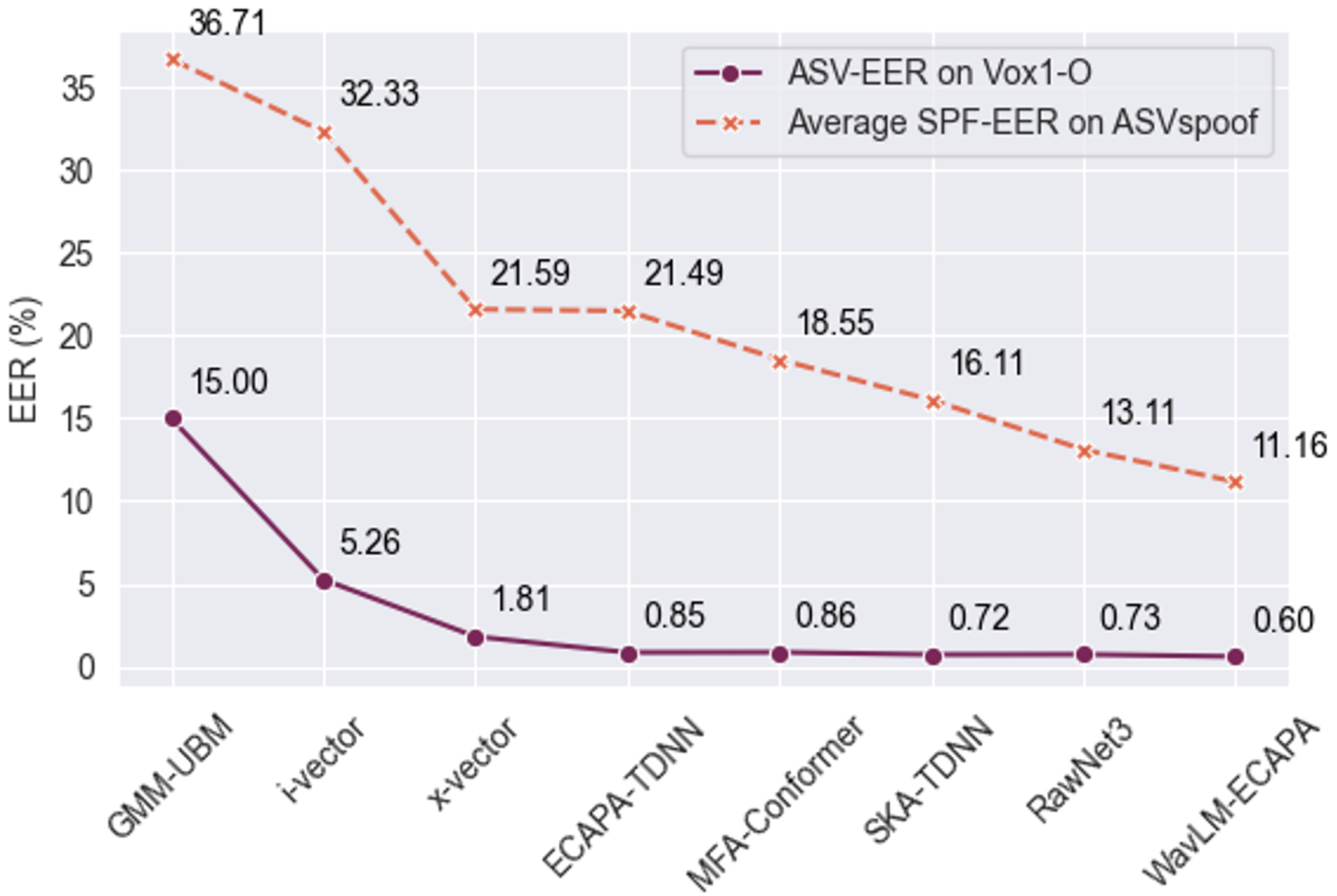}
    \caption{
    Average Spoof Equal Error Rates (SPF-EERs) on 29 different spoofing attacks, chronologically displayed using eight automatic speaker verification (ASV) systems.
    The SPF-EER adopts spoof trials in place of conventional non-target trials, where in a spoof trial, the test utterance is a system-generated voice of the target speaker.
    Conventional EERs (equivalent to SV-EER in \cite{jung2022sasv}) of the ASV systems in the Vox1-O evaluation protocol are also reported as a reference. 
    }
    \label{fig:asv_progression_2019}
    \vspace{-19pt}
\end{figure}

In response, a specialized area of study has emerged aimed at enhancing ASV systems with integrated spoofing detection capabilities, thus introducing a third type of trial: the spoof trial~\cite{jung2022sasv,mun2023towards,torgashov2023id,wu2015asvspoof}. Despite the addition, these extended systems, termed Spoofing-robust ASV (SASV), retain a binary classification function, authenticating target trials while rejecting all others. Initial efforts in SASV development combined separately developed ASV and countermeasure (CM) subsystems~\cite{ge2023can,liu2024generalizing,shim2022baseline,wu2015asvspoof}.\footnote{The CM system does not require a target enrolled speaker. It makes a binary decision on whether the input is from humans or synthesized.}  More recent approaches have explored integrated solutions~\cite{mun2023towards,shim2020integrated,todisco2018integrated}, utilizing the adaptive power of neural networks to assess both the identity of the speaker and the authenticity of the speech concurrently, aligning with the singular goal of accepting target trials.

While ASV systems are generally vulnerable to spoofing, not all attacks are equally effective. The ASVspoof 2019 \cite{wang2020asvspoof} revealed that different attacks impact the ASV Equal Error Rate (EER) to varying extents. 
Further investigation \cite{ge2023can} has uncovered instances where an ASV system can reject spoofing attacks, even when their training regimes did not explicitly account for such scenarios (i.e., {\em zero-shot} in terms of data and scenario).
This finding indicates that ASV systems possess an inherent ability to reject spoof attempts, especially when the imitation of the target speaker's characteristics falls short.
This insight leads to two key questions: (1) Can future enhancements in ASV technology, focused exclusively on refining speaker discrimination as currently practiced, prove robust enough to counter both present and future spoofing challenges? (2) How effectively can current ASV systems, utilizing a zero-shot strategy, repel spoofing attacks, and which precise varieties of attacks are they adept at blocking?

To address the first question, we present a chronological analysis examining the correlation between advancements in ASV technology and their inherent defense against spoofing attacks. 
Our study spans a broad spectrum of ASV technologies, ranging from the Gaussian Mixture Model-Universal Background Model (GMM-UBM)~\cite{reynolds2000speaker} developed in the pre-deep learning era to the latest Self-Supervised Learning (SSL) representation-based Enhanced Deep Speaker Embedding (ECAPA-TDNN~\cite{desplanques2020ecapa}).
Illustrated in Figure~\ref{fig:asv_progression_2019} and detailed further in Section~\ref{sec:results}, our findings reveal a clear evolution in ASV systems' natural defense mechanisms against spoofing attacks.\footnote{This observation, while somewhat expected -- since improved speaker discrimination inherently complicates successful spoofing attempts -- represents the first documentation of a clear evolutionary trend.} 
Notably, with conventional ASV error rates plummeting from $13.00$\% to 0.60\%, the systems' robustness to spoofing has increased fourfold, although these ASV systems were never explicitly designed to counter spoof scenarios.
This demonstrates a remarkable {\em zero-shot} capability; the ASV systems effectively countered spoofing attacks without any prior training on such scenarios or data. However, it's important to note that the current EER, above 10\%, remains too high for practical application in real-world scenarios.

To address the second question, we conduct a parallel chronological analysis focused on the evolution of attack technologies. Our examination of 29 different TTS/VC systems reveals a consistent pattern; speech data spoofed using the latest algorithms proves significantly more adept at fooling ASV systems. This outcome reflects the continuous advancement in spoofing techniques.

Our analysis leads to two critical insights: (1) advancements in ASV are enhancing the robustness of existing spoofing attacks; (2) progress in TTS/VC is far outpacing that in ASV.
Given this dynamic, we emphasize the urgent need for the ASV community to intensify efforts in crafting SASV systems.\footnote{We envision a future where the term ASV inherently signifies SASV, recognizing spoofing-robustness as a fundamental requirement for any ASV system to effectively serve as an authentication mechanism.} Whether through integrating ASV and CM subsystems or by developing single neural networks, the enhancement of spoofing robustness is paramount. Without it, ASV systems lack the necessary robustness to serve as reliable authentication mechanisms.

\section{ASV systems}
\label{sec:asv_systems}

Given the limited space, we briefly outline the eight ASV systems under investigation in this work.
RawNet3~\cite{jung2022pushing} is the only system not relying on handcrafted acoustic features such as mel-spectrograms for input. Refer to the references for details.

\newpara{GMM-UBM} utilises a generative modeling approach for ASV~\cite{reynolds2000speaker,larcher2014text,kinnunen2016utterance}. 
It trains a GMM with the training set to form a Universal Background Model (UBM). 
The GMM-UBM system evaluates the log-likelihood ratio between the GMM adapted toward the target speaker and the UBM for each input.

\newpara{i-vector.}
\cite{dehak2010front} captures speaker and channel variabilities in a unified low-dimensional space. It projects a GMM supervector, derived from a UBM, into this compact latent space using a total variability matrix.

\newpara{x-vector.}
\cite{snyder2018x} leverages time-delay neural networks (TDNNs) for frame-level analysis, employing varying dilation rates.  
A statistics pooling layer then aggregates these frame embeddings into a singular utterance-level embedding through mean and standard deviation calculations, followed by two dense layers for further processing.  

\newpara{ECAPA-TDNN}
\cite{desplanques2020ecapa} employs Res2Net as the backbone processing block with a squeeze-excitation module for frame-level processing. Additional techniques, such as multi-layer feature aggregation, are adopted. Context- and channel-dependent statistics pooling is devised to aggregate frame embeddings into utterance-level embedding.

\newpara{MFA-Conformer}
\cite{zhang2022mfa} combines CNN sub-sampling and Conformer encoder layers for frame embedding. An attentive statistics pooling layer combines frame embeddings into an utterance embedding. This process involves concatenating and normalizing frame embeddings from all encoder layer outputs through layer normalisation~\cite{ba2016layer}.\footnote{In our implementation, we adopt the context- and channel-dependent statistics pooling from ECAPA-TDNN.} 

\newpara{SKA-TDNN.}
Building upon ECAPA-TDNN, SKA-TDNN~\cite{mun2023frequency} introduces a novel frame embedding processing mechanism (``msSKA block") and a feature-enhancing module ("fcwSKA block") employing selective kernel attention. This is followed by context- and channel-dependent pooling, batch normalization, and dense layers for utterance-level integration.

\newpara{RawNet3}
\cite{jung2022pushing} is a CNN-based system optimized for processing raw waveforms directly, employing an analytic filterbank~\cite{pariente2020filterbank} layer as a substitute for traditional acoustic features. Its modeling approach for both frame and utterance levels parallels ECAPA-TDNN, with key adjustments for raw waveform input compatibility. These modifications include strategically using pooling layers to condense the sequence length. 

\newpara{WavLM-Large with ECAPA-TDNN.}
This system combines WavLM-Large~\cite{chen2022wavlm}, an SSL model with strong representations for several downstream tasks, including ASV, with ECAPA-TDNN.
We freeze the WavLM and use its representation as a feature.
ECAPA-TDNN is trained using the WavLM representations in place of conventional acoustic features. 

\section{Corpora}

\subsection{Training}
For training our ASV systems, except for the GMM-UBM, we utilize the VoxCeleb 1 and 2 corpora~\cite{nagrani2017voxceleb,chung2018voxceleb2}, which feature celebrity utterances sourced from YouTube. These datasets are the preferred choice for recent ASV research.
The combined training set encompasses over 2.7k hours of speech, derived from 1.2 million utterances across 7.3k speakers.
The WavLM-Large~\cite{chen2022wavlm} we use was pre-trained with external corpora.
We train two GMM-UBMs, utilizing the training partitions of ASVspoof 2015 and ASVspoof 2019 logical access, respectively.

\subsection{Evaluation}
We select the ASVspoof 2015~\cite{wu2015asvspoof} and 2019 logical access (LA)~\cite{wang2020asvspoof} corpora to assess our ASV systems' performances when faced with spoofing attacks. Other corpora were excluded due to the lack of an official SASV evaluation protocol, which is crucial for our analysis.\footnote{Most corpora have evaluation protocols for assessing anti-spoofing systems where a binary decision of ``bonafide'' and ``spoof'' needs to be made. For the analysis in this work, where ASV systems are evaluated in a zero-shot manner with spoof attacks, we need an SASV evaluation protocol, where input is a pair of an enrollment speaker and a test utterance. Each trial belongs to one of the three categories, ``target'', ``non-target'', and ``spoof''.} Table~\ref{tab:asvspoof_corpora} describes the statistics of the two corpora. Table~\ref{tab:attack_spec} provides detailed descriptions of each spoofing attack algorithm, categorizing them as follows: ``Group 1" excludes neural networks, ``Group 2" incorporates neural networks solely in the acoustic model, ``Group 3" utilizes neural networks in both acoustic and language models, and ``Group 4" consists of non-parametric systems.

\begin{table}
  \caption{Statistics of the ASVspoof 2015's evaluation partition and 2019 LA's development and evaluation partitions~\cite{wu2015asvspoof,wang2020asvspoof}.}
  \label{tab:asvspoof_corpora}
  \resizebox{\columnwidth}{!}{
  \begin{tabular}{lccc}
  \toprule
    Corpus & \# Speaker & \# Utterance & \# Algorithms\\
  \toprule
  ASVspoof 2015 & $46$ & $193,404$ & $10$\\
  ASVspoof 2019 LA & $87$ & $96,081$ & $19$\\
  \bottomrule
  \end{tabular}
  }
  \vspace{-10pt}
\end{table}

\newpara{ASVspoof 2015.} This corpus, pioneering in the realm of spoofing-robust ASV, includes an evaluation set with ten spoofing attacks derived from TTS and VC technologies, as detailed in Table~\ref{tab:attack_spec} with IDs beginning with ``S."

\newpara{ASVspoof 2019 logical access} is the most widely used corpora for developing spoofing-robust ASV and anti-spoofing systems. Its development and evaluation partitions comprise 19 spoofing attacks. Details on the attacks are presented Table~\ref{tab:attack_spec} with ``ID'' starting with ``A.'' 

\newpara{Vox1-O.} Adopted as the benchmark protocol, this involves data from 40 speakers in the VoxCeleb1 test set. We adopt the protocol to show our ASV systems' performance in Figure~\ref{fig:asv_progression_2019}.

\begin{table}[t!]
  \caption{
  Spoofing attacks in ASVspoof 2015 (S*) and 2019 LA corpora (A*). They are grouped based on acoustic and waveform models, and A18 to A10 are sorted based on when the system was proposed. 
  DNN-based \textcolor{blue}{acoustic} and \textcolor{red}{waveform} models are highlighted in the second and third columns, respectively. 
  }
  \label{tab:attack_spec}
  \vspace{-10pt}
 \centering
 \resizebox{0.8\columnwidth}{!}{%
  \begin{tabular}{rrrrr}
  \toprule
  Group & ID & Type & \shortstack{Acoustic model} &  \shortstack{Waveform model} \\
  \toprule
\multirow{12}{*}{1} & \cellcolor{color_vc}A18&VC&\cellcolor{color_am_sm}i-vector + PLDA&\cellcolor{color_wm_dsp}LPC\\
& \cellcolor{color_vc}S5&VC&\cellcolor{color_am_sm}GMM&\cellcolor{color_wm_dsp}MLSA\\
& \cellcolor{color_vc}A06&VC&\cellcolor{color_am_sm}GMM&\cellcolor{color_wm_dsp}spectral filtering\\
& \cellcolor{color_vc}A19&VC&\cellcolor{color_am_sm}GMM&\cellcolor{color_wm_dsp}spectral filtering\\
& \cellcolor{color_vc}S2&VC&\cellcolor{color_am_sm}Linear reg.&\cellcolor{color_wm_dsp}STRAIGHT\\
& \cellcolor{color_vc}S1&VC&\cellcolor{color_am_sm}DTW&\cellcolor{color_wm_dsp}STRAIGHT\\
& \cellcolor{color_vc}S6&VC&\cellcolor{color_am_sm}GMM + GV&\cellcolor{color_wm_dsp}STRAIGHT\\
& \cellcolor{color_vc}S7&VC&\cellcolor{color_am_sm}GMM + GV&\cellcolor{color_wm_dsp}STRAIGHT\\
& S3&TTS&\cellcolor{color_am_sm}NLP + HMM&\cellcolor{color_wm_dsp}STRAIGHT\\
& S4&TTS&\cellcolor{color_am_sm}NLP + HMM&\cellcolor{color_wm_dsp}STRAIGHT\\
& \cellcolor{color_vc}S8&VC&\cellcolor{color_am_sm}GMM-tensor&\cellcolor{color_wm_dsp}STRAIGHT\\
& \cellcolor{color_vc}S9&VC&\cellcolor{color_am_sm}DTW + Kernel reg.&\cellcolor{color_wm_dsp}STRAIGHT\\
\midrule
\multirow{9}{*}{2} & \cellcolor{color_vc}A05&VC&\cellcolor{color_am_dnn}VAE&\cellcolor{color_wm_dsp}WORLD\\
& \cellcolor{color_vc}A17&VC&\cellcolor{color_am_dnn}VAE&\cellcolor{color_wm_dsp}waveform filtering\\
& A13&TTS&\cellcolor{color_am_dnn}TTS + VC(DNN)&\cellcolor{color_wm_dsp}waveform filtering\\
& A09&TTS&\cellcolor{color_am_dnn}NLP + RNN&\cellcolor{color_wm_dsp}Vocaine\\
& A14&TTS&\cellcolor{color_am_dnn}TTS + VC(DNN)&\cellcolor{color_wm_dsp}STRAIGHT\\
& A03&TTS&\cellcolor{color_am_dnn}NLP + DNN&\cellcolor{color_wm_dsp}WORLD\\
& A02&TTS&\cellcolor{color_am_dnn}NLP + HMM-DNN&\cellcolor{color_wm_dsp}WORLD\\
& A07&TTS&\cellcolor{color_am_dnn}NLP + RNN-GAN&\cellcolor{color_wm_dsp}WORLD\\
& A11&TTS&\cellcolor{color_am_dnn}DNN(end2end)&\cellcolor{color_wm_dsp}Griffin-Lim\\
\midrule
\multirow{5}{*}{3} & A08&TTS&\cellcolor{color_am_dnn}NLP + HMM-DNN&\cellcolor{color_wm_dnn}Dilated CNN\\
& A01&TTS&\cellcolor{color_am_dnn}NLP + HMM-DNN&\cellcolor{color_wm_dnn}WaveNet\\
& A12&TTS&\cellcolor{color_am_dnn}NLP + RNN&\cellcolor{color_wm_dnn}WaveNet\\
& A15&TTS&\cellcolor{color_am_dnn}TTS + VC(DNN)&\cellcolor{color_wm_dnn}WaveNet\\
& A10&TTS&\cellcolor{color_am_dnn}DNN(end2end)&\cellcolor{color_wm_dnn}WaveRNN\\
\midrule
\multirow{3}{*}{4} & S10&\multirow{3}{*}{TTS}&\multirow{3}{*}{\cellcolor{color_am_sm}NLP + Unit-selection}&\multirow{3}{*}{\cellcolor{color_wm_dsp}Waveform concat.}\\
& A04 \\
& A16 \\
\bottomrule
  \end{tabular}
  }%
  \vspace{-20pt}
\end{table}

\begin{figure*}
    \centering
    \includegraphics[width=\textwidth]{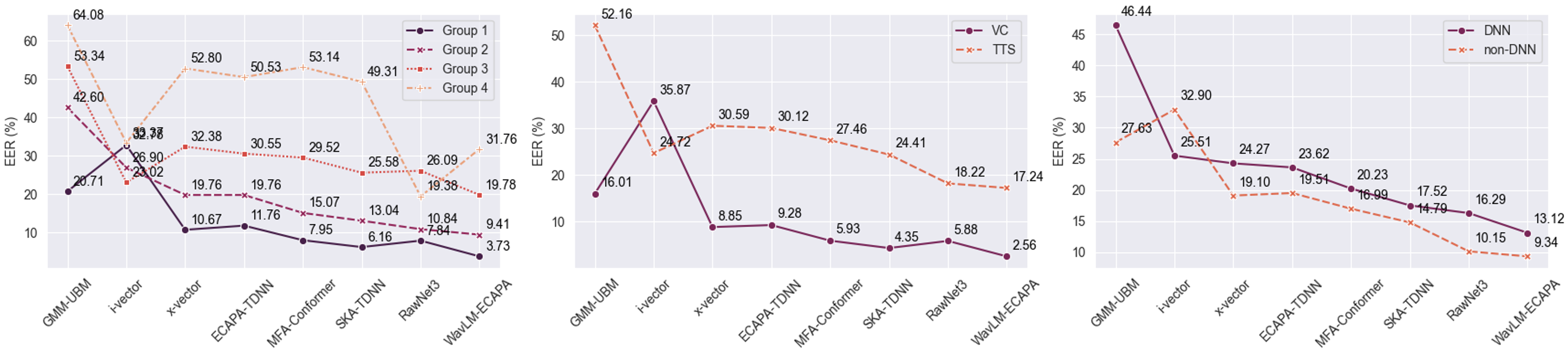}
    \vspace{-15pt}
    \caption{
    Detailed analyses on chronologically sorted eight ASV systems. (left): different groups of spoofing attacks. (middle): TTS vs. VC attacks. (right) DNN vs. non-DNN-based spoofing attacks. Group 1: does not involve neural networks. Group 2: only the acoustic model is a neural network. Group 3: acoustic and waveform models both are neural networks. Group 4: non-parametric systems.
    }
    \label{fig:detailed_asv_chronological}
    \vspace{-10pt}
\end{figure*}
\begin{figure*}
    \centering
    \includegraphics[width=\textwidth]{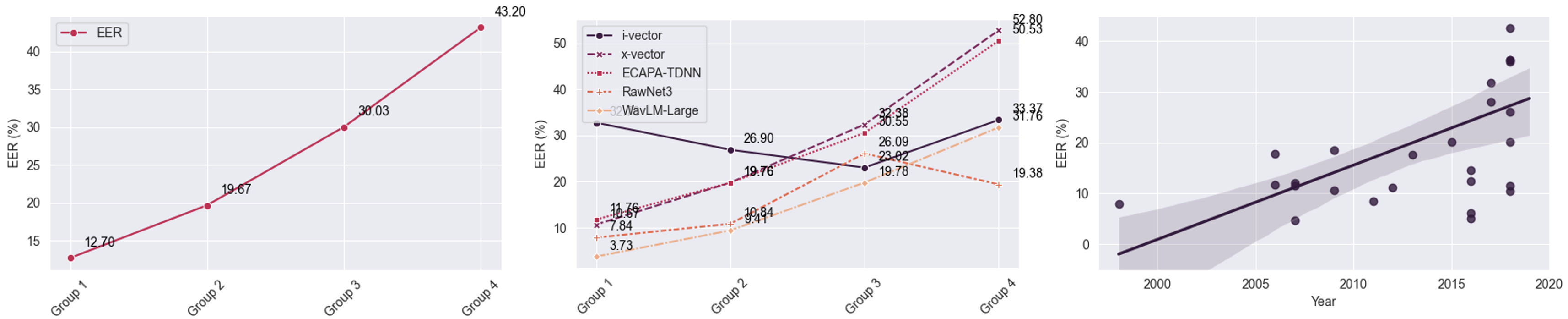}
    \vspace{-20pt}
    \caption{
    Analyses from the perspective of spoofing attacks. 
    (left): averaged ASV results in terms of groups. (middle): five ASV systems' results in terms of groups.
    (right): averaged ASV results of all 29 spoofing attacks in terms of years.
    }
    \label{fig:cm_perspective}
    \vspace{-15pt}
\end{figure*}

\section{Experimental configurations}
\label{sec:exp_cfg}
\subsection{GMM-UBM and i-vector}

\newpara{GMM-UBM.}
We employ three GMM-UBM systems in this work.
GMM-UBM performance in Figure~\ref{fig:asv_progression_2019} is from \cite{nagrani2017voxceleb}.
We trained two GMM-UBM systems using the MSR Identity toolkit~\cite{sadjadi2013msr}.
These UBMs were trained on bonafide training data from ASVspoof 2015 and 2019 LA, excluding spoofed utterances. 
Each system was employed to assess the respective evaluation set.\footnote{This decision aimed to expose GMM-UBM systems to a lesser domain mismatch compared to the other seven ASV systems, attempting to compensate for GMM-UBM's relatively lower performance. Despite this advantage, GMM-UBM remains the least effective.}

\newpara{i-vector.}
We use the Kaldi toolkit~\cite{Kaldi-Povey2011} to train the i-vector system. Our implementation follows the original recipe with a modification: we omit the VoxCeleb2 test set from the training dataset to ensure consistency in training data across all ASV systems.

\vspace{-0.2cm}
\subsection{DNN-based systems}
\vspace{-0.2cm}
For the remaining six DNN-based systems, we leveraged publicly available pre-trained systems from ESPnet-SPK~\cite{watanabe2018espnet,jung2024espnet}.
Our methodology aimed to minimize hyper-parameter optimization, striving for consistency in configuration settings across these systems to mitigate the influence of irrelevant variables. 
Unless mentioned otherwise, the below configuration applies to all six systems.
We set the batch size to 512 and trained the model for 40 epochs.
Models use Adam optimizer~\cite{kingma2015adam} for training. 
A cosine annealing learning rate scheduler with warm-ups and restarts schedules the learning rate with the peak of $1e^{-3}$ and the minimum of $5e^{-6}$. 
We also augmented the data during training with Musan~\cite{snyder2015musan} noise and RIR~\cite{ko2017study} reverberations.
80-dimensional mel-spectrograms with a 25ms window and a 10ms shift size are used for the acoustic feature.
Speaker embeddings have 192 dimensions.
Training objective functions comprise AAM-softmax~\cite{deng2019arcface} with sub-center and inter top-k losses~\cite{zhao2021speakin}.
Details on hyper-parameters and training schemes are available at \texttt{\url{https://github.com/espnet/espnet/blob/master/egs2/voxceleb/spk1/README.md}}.

\noindent\textbf{x-vector} implementation follows the original architecture proposed in \cite{snyder2018x}. Its speaker embeddings are 512-dimensional. \textbf{MFA-Conformer} adopts six encoder blocks, each with 512-dimensional output and a CNN kernel size of 15. We apply a longer warm-up of 10k steps. \textbf{WavLM-Large with ECAPA-TDNN}
uses the weighted summation of the WavLM-Large model's layer-wise outputs.

\vspace{-0.2cm}
\subsection{Metrics}
\vspace{-0.2cm}
All outcomes are assessed using two evaluation metrics. The conventional EER for ASV involves target and non-target trials. The Spoofing Equal Error Rate (SPF-EER) is determined using target trials and spoofed non-target trials, evaluating the ASV system's susceptibility when non-target trials are substituted with spoofing trials. Refer to \cite{jung2022sasv,shim2024adcf} for details.

\vspace{-0.3cm}
\section{Results}
\label{sec:results}

\subsection{ASV performances}
Figure~\ref{fig:asv_progression_2019} presents the EERs of the eight ASV systems on the Vox1-O dataset (dotted line) alongside their SPF-EER across 29 spoofing attacks in the ASVspoof 2015 and 2019 LA corpora. 
Performances on Vox1-O demonstrate the evolution of conventional ASV, while performances on ASVspoof highlight ASV systems' zero-shot (i.e., effortless) capability against out-of-domain spoofing attacks.

\subsection{Does ASV progression naturally enhance defense against spoofing attacks?}
Analysis from Figure~\ref{fig:asv_progression_2019} demonstrates a significant improvement in ASV performance, with EER reducing from 36.71\% in the GMM-UBM system to 11.16\% in the WavLM-Large with ECAPA-TDNN. Figure~\ref{fig:detailed_asv_chronological} offers an in-depth analysis of this progression. 
The left section of this figure indicates a general trend of increasing EER across Groups 1 to 4,\footnote{Group 1 represents the earliest systems without neural networks, while Group 3 includes the most recent systems employing neural networks for both acoustic and waveform modeling.} suggesting that advancements in ASV technology do not inherently confer enhanced defenses against increasingly sophisticated spoofing attacks. 
The right section confirms this observation, showing that EERs for spoofing attacks utilizing neural network-based TTS and VC systems remain consistently more challenging than their counterparts. 
Intriguingly, the middle section reveals that VC attacks consistently yield lower EERs than TTS attacks, highlighting a distinct vulnerability pattern within ASV systems' response to different spoofing methodologies.

\begin{table}
  \caption{State-of-the-art SASV systems.}
  \vspace{-5pt}
  \centering
  \label{tab:sasv_sota}
  \resizebox{0.7\columnwidth}{!}{
  \begin{tabular}{lcc}
  \toprule
    System & ASV-EER & SPF-EER\\
  \toprule
  Jointly optimised~\cite{Alenin2022A} & 0.11 & 0.17\\
  Single neural network~\cite{mun2023towards} & 1.27 & 1.23\\
  \bottomrule
  \end{tabular}
  }
  \vspace{-15pt}
\end{table}

\vspace{-5pt}
\subsection{Insights from the spoofing attacks}
Figure~\ref{fig:cm_perspective} offers a viewpoint from the perspective of spoofing attacks. The left-side plot shows a trend of increasing EERs chronologically across Groups 1 to 3. We also find that unit-selection-based non-parametric TTS systems, group 4, pose the most significant challenge. 
The middle plot examines the performance of the five individual ASV systems across different groups to identify any distinct trends. Interestingly, neural network-based ASV systems exhibit similar patterns of monotonic increase, whereas the i-vector system demonstrates generalized EER performances across various groups of attacks, albeit with generally poor performance.
Finally, the right-side plot arranges the 29 attacks according to the systems' devised years, highlighting the consistent upward trajectory of EERs. This consistent increase across different timeframes underscores the growing necessity for research into spoofing-resistant ASV technologies, specifically SASV systems.

\subsection{Discussion: potential of SASV research}
In closing, we demonstrate the bright potential of SASV research through a snapshot of the current state-of-the-art SASV systems~\cite{Alenin2022A,mun2023towards} in Table~\ref{tab:sasv_sota}. The results are benchmarked using the ASVspoof 2019 LA corpus in the way it has been utilized in the SASV 2022 challenge; they are SASV systems rejecting non-target and spoof trials, not CM systems making a binary decision of bonafide and spoof.
It's important to note that the evaluations consider only 13 of the attacks, making a direct comparison with other results in this paper incompatible. 
Despite the $13$ analyzed attacks being more complex than the $16$ not examined, the results reveal that current SASV systems (1.23\% SPF-EER) significantly outperform ASV systems (11.16\% SPF-EER) in dismissing spoof trials while also accurately rejecting non-target trials.
This outcome highlights the promising potential and evolving capability of SASV systems in the ongoing battle against spoofing threats.

\section{Conclusion and future works}
Our investigation centered on whether the evolution of contemporary ASV systems, which traditionally do not account for spoofing scenarios, inherently develops defenses against such attacks. Through comprehensive analysis, we have confirmed a gradual improvement in ASV systems' ability to reject spoofed inputs. However, this progress lags significantly behind the rapid advancements in speech-generation technologies.
Our analyses call for an urgent push within the ASV community towards developing either integrated ASV and CM subsystems or adopting single, unified SASV neural network approaches. Given the pace of advancements in deep learning, we strongly advocate for a focused effort on the latter, anticipating it will yield significant breakthroughs in enhancing ASV's resilience against spoofing attacks.

\clearpage
\section{Acknowledgement}
Experiments of this work used the Bridges2 system at PSC and Delta system at NCSA through allocations CIS210014 and IRI120008P from the Advanced Cyberinfrastructure Coordination Ecosystem: Services \& Support (ACCESS) program, supported by National Science Foundation grants \#2138259,\#2138286, \#2138307, \#2137603, \#2138296.
This study is partially supported by JST AIP Acceleration Research (JPMJCR24U3), JST PRESTO (JPMJPR23P9), Japan, and by the ANR Bruel project (ANR-22-CE39-0009), France, and by ONR N00014-231-2086, USA.

\bibliographystyle{IEEEtran}
\bibliography{shortstrings,mybib}


\end{document}